\documentclass[aip, amsmath,amssymb,reprint]{revtex4-1}
\usepackage{graphicx}
\usepackage{dcolumn}
\usepackage{bm}
\usepackage[mathlines]{lineno}
\usepackage[utf8]{inputenc}
\usepackage[T1]{fontenc}
\usepackage{mathptmx}
\usepackage{mathtools}
\usepackage{amsmath}
\usepackage{etoolbox}
\usepackage{subfigure}
\usepackage{epstopdf}

\usepackage[bookmarksnumbered=true,bookmarksopen=true]{hyperref}
\hypersetup{colorlinks=true,%
	        linkcolor =blue, %
        	citecolor =blue, %
	        urlcolor  =blue }

\makeatletter
\def\@email#1#2{
 \endgroup
 \patchcmd{\titleblock@produce}
  {\frontmatter@RRAPformat}
  {\frontmatter@RRAPformat{\produce@RRAP{*#1\href{mailto:#2}{#2}}}\frontmatter@RRAPformat}
  {}{}}
\makeatother
\allowdisplaybreaks

\begin{document}
\preprint{AIP/123-QED}

\title[Multi-Radical Lipkin-Meshkov-Glick Model for Avian Navigation]
{Multi-Radical Lipkin-Meshkov-Glick Model for Avian Navigation}
\author{Jia-Yi Wu}
\author{Xin-Yuan Hu}
\author{Hai-Yuan Zhu}
\author{Ru-Qiong Deng}
\author{Qing Ai}
 \email{aiqing@bnu.edu.cn}
 \homepage{http://quanphys.bnu.edu.cn}
\affiliation{%
Department of Physics, Applied Optics Beijing Area Major Laboratory, Beijing Normal University, Beijing 100875, China
}%

\date{\today}

\begin{abstract}
  The mechanism of avian navigation is an important question for global scientists. One of the most famous candidates is the radical pair mechanism (RPM), which shows that avian navigation is achieved by detecting the amount of products from chemical reactions, which exhibit different properties for singlet and triplet spin states. Based on the theory, we explore the amount of the the singlet recombination product by changing the coupling strength and direction of the magnetic field. The radical pair in our model is coupled with a Lipkin-Meshkov-Glick (LMG) bath, which proposes  attractive results at quantum phase transition (QPT) point, especially when the magnetic field is parallel with the original direction of the pair. We find weakest navigation ability in QPT point according to one of our proposing methods. And we find invariant subspace in our Hamiltonian when the magnetic field is vertical to the plane of the radical pair and the bath, which will greatly speed up our calculation. Our results could help the exploration of the design principle of artificial avian navigation.
\end{abstract}

\maketitle
\section{Introduction}\label{sec:Introduction}
In the seminal book, \textit{What's Life?}, Schr\"{o}digner presented the idea that physical laws in macroscale is decided by statistical behaviors in microscale \cite{WhatsLife}. Ever since then, physical mechanisms have been applied to explain various physiological processes in living beings, e.g. coherent energy transfer in natural photosynthesis \cite{Lambert2013NP,Cao2020SA,Tao2020SciBull,Rey2013JPCL,Ai2013JPCL,Collini2010Nat}, a quantum model for protein folding \cite{Mao2021scicn}, and the radical-pair mechanism for migrating birds \cite{Ritz2000BJ,Solovyov2008PSB}.

Among the 50 species, which include insects, crustaceans, fish, amphibians, reptiles, mammals and birds, and utilize geomagnetic field for navigation, migratory birds have been intensively studied \cite{Mouritsen2018Nat}. In the radical-pair hypothesis, the radical pair in the cryptochrome is activated by photons. The anisotropic magnetic field due to the nearby nuclear spins can result in the reaction yield which is sensitive to orientation of the geomagnetic field. The cryptochrome is located at the avian retina and the visual modulation patterns can be influenced by the geomagnetic field through the above spin-selective chemical reaction \cite{Ritz2000BJ,Solovyov2008PSB}. In a model system, carotenoid–porphyrin–fullerene, the radical pair is photochemically activated. When the probe light is not perpendicular to the simulated geomagnetic field, the lifetime of the radical pair in chemical reaction \textit{in vitro} shows clear dependence on the orientation of the magnetic field \cite{Maeda2008Nat}. \cite{Xu2021Nat} Such kind of spin-dependent chemical reactions can be well described by the Holstein model which is generalized to naturally include spin degree of freedom of the radicals \cite{Yang2012PRA}. Inspired by these interesting discoveries, bionic quantum coherent devices for navigation have been proposed, e.g. criticality-enhanced quantum compass \cite{Cai2012PRA}.

We notice that nuclear spins play a crucial role in the proposed radical-pair mechanism, because the homogeneous geomagnetic field can not induce the transition between the initial singlet state and the triplet states \cite{Ritz2000BJ,Yang2012PRA}. Recent studies have suggested that the hyperfine interaction is not necessary for the magnetoreception, and dipolar interactions between radicals beyond the radical-pair model can also assist navigation by the magnetic-field-sensitive chemical reactions \cite{Keens2018PRL}. Researchers also find that non-uniform magnetic field can also be used to radical pair mechanism so as to navigate \cite{Cai2011PRL}. It is also shown that the QPT in the nuclear spins described by the Ising model in the transverse field can improve the sensitivity for navigation \cite{Cai2012PRA}. Considering the above discoveries, it is quite natural to ask: Can we use the QPT in the multi-radical model to optimize navigation? Can the QPT beyond the Ising model in the transverse field be used, e.g. the Lipkin-Meshkov-Glick (LMG) model?

The LMG model was proposed to offer exact suctions or solutions with ordinary perturbation to quantum many-body problems \cite{Lipkin1965NuclPhys,Meshkov1965NuclPhys,Glick1965NuclPhys}. It describes a system with $N$ fermions distributed on two $N$-fold degenerated energy levels. Different from the Ising model in the transverse field or $XY$ model which only consider the nearest-neighbor interactions \cite{PRLGarnerone2009,PRACarmen2008,Ai2008PRA}, the LMG model take all interactions of the spins into consideration based on the mean-field theory \cite{Salvatori2014PRA,Quan2007PRA}. Analytical methods have been use to obtain eigen values and corresponding eigenfunctions for LMG model \cite{PAN19991PLB}, or the low-energy spectrum and associated generator matrix elements in the eigen energy basis of LMG  \cite{Rosensteel2007JPAMT}. An interesting result that the QPT also exists in the LMG model \cite{Quan2007PRA,Octavio2006PRB}, inspires us to improve the sensitivity of magnetic detection by the QPT in the multi-radical which is described by the LMG model.

This paper is organized as follows: In Section~\ref{Radical Pair Models}, we propose two new model to explain avian navigation based on the LMG model with many radicals. In Section~\ref{section3}, we are going to explore the vertical situation. At last, the main conclusions are summarized in Section~\ref{Conclusion} and we also discuss the possibilities of applications of our discoveries.

\section{Multi-Radical Models for Avian Navigation}\label{Radical Pair Models}
\begin{figure}[t]
\includegraphics[width=7.5cm]{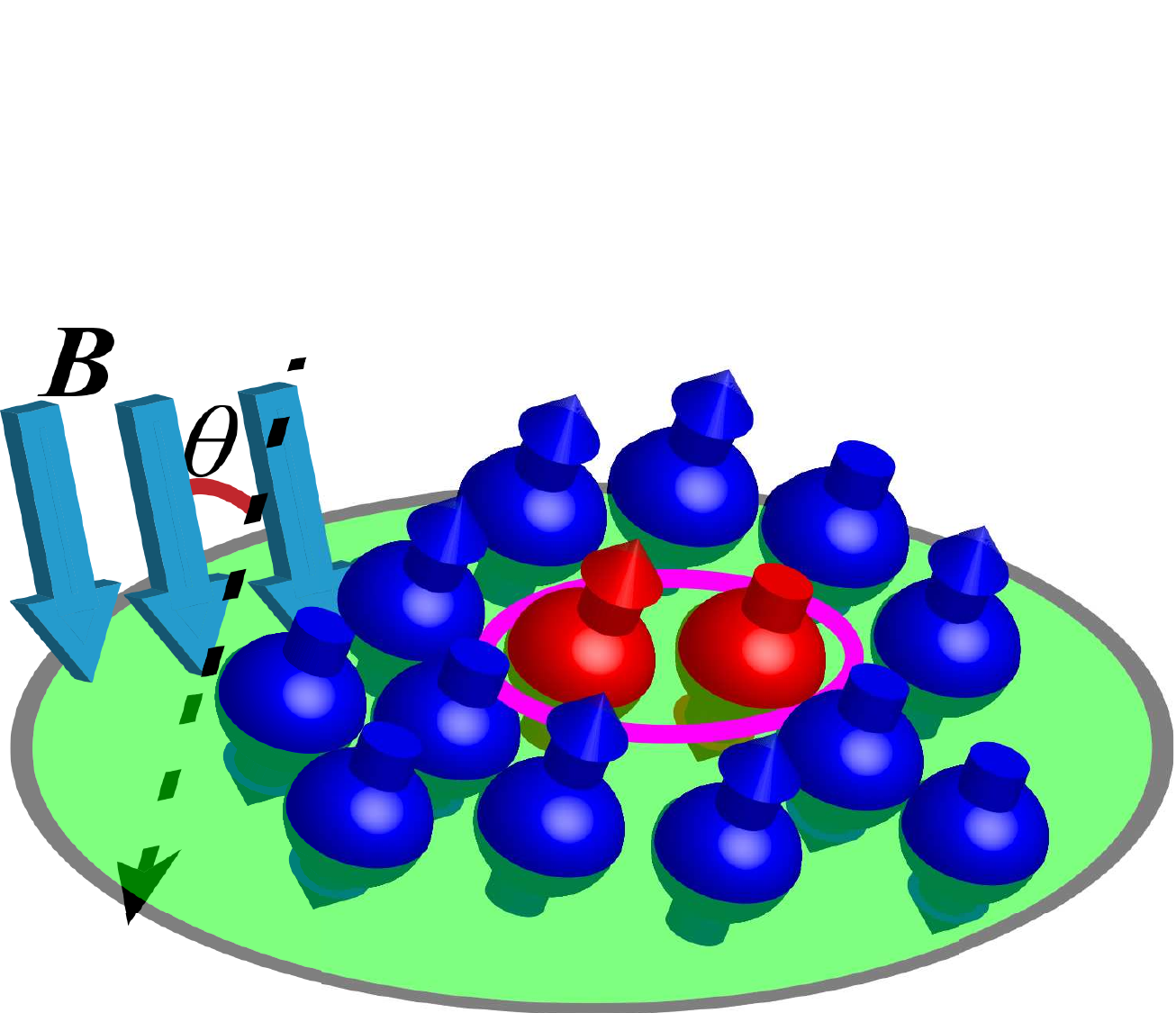}
\caption{Schematic of the radical-pair system interacting a multi-radical bath. The red spheres represent the radical pair, while the blue spheres are the radicals in the bath described by the LMG model. The angle between the geomagnetic field and the normal of the plane is $\theta$.}
\label{fig:scheme}
\end{figure}

As shown in Fig.~\ref{fig:scheme}, we consider a radical pair system coupling to $N$ radicals which are described by the LMG model. The geomagnetic field is applied along the direction $\cos\theta \vec{e}_z+\sin\theta \vec{e}_x$. The Hamiltonian of the total system reads
\begin{equation}
	\begin{aligned}
		H=&-\frac{\lambda}{N}\sum_{i<j}^{N}\left(\sigma_{i}^{x}\sigma_{j}^{x}+\sigma_{i}^{y}\sigma_{j}^{y}\right)-\lambda^{\prime}\left(\sigma_{a}^{x}\sigma_{b}^{x}+\sigma_{a}^{y}\sigma_{b}^{y}\right)\\
		&-{\lambda}_{1}\sum_{i=1}^{N}\left[\sigma_{i}^{x}\left(\sigma_{a}^{x}+\sigma_{b}^{x}\right)+\sigma_{i}^{y}\left(\sigma_{a}^{y}+\sigma_{b}^{y}\right)\right]\\
		&-\sum_{i=1}^{N}\left(\cos{\theta}\sigma_{i}^{z}+\sin{\theta}\sigma_{i}^{x}\right)\\
		&-\cos{\theta}\left(\sigma_{a}^{z}+\sigma_{b}^{z}\right)-\sin{\theta}\left(\sigma_{a}^{x}+\sigma_{b}^{x}\right),
		\label{eq: H0}
	\end{aligned}
\end{equation}
where $\sigma_{i}^{\alpha}$ ($i=1,2,...,N,~\alpha=x,y,z$) represent the Pauli matrices of the radical bath, $\sigma_{a}^{\alpha}$ and $\sigma_{b}^{\alpha}$ are the Pauli matrices of the radical pair. $\lambda/N$ is the interaction between the radicals in the bath, which is inversely proportional to the total number of the spins according to the mean field theory. Here, we assume that the coupling strength between the radical pair system and the bath, $\lambda_{1}$, can be approximated to the interaction of bath, $\lambda/N$. $\lambda^{\prime}$ is the interaction between the two radicals, which is $\lambda/2$ according to the LMG model. Notice that when the direction of the geomagnetic field is reversed, i.e., $\theta\rightarrow\pi-\theta$, avian navigation will not be influenced \cite{Wolfgang1972Science,Ritz2000BJ}.

We define the collective spin operators for the bath and radical as, $J_{N}^{\alpha}=\frac{1}{2}\sum_{i=1}^{N}\sigma_{i}^{\alpha}$ and   $S^{\alpha}=\frac{1}{2}\left(\sigma_{a}^{\alpha}+\sigma_{b}^{\alpha}\right)$ ($x,y,z$). Correspondingly, the raising and lowering operators are given as $J_{N}^{\pm}=J_{N}^{x}{\pm} \mathrm{i} J_{N}^{y}$ and $S^{\pm}=S^{x}{\pm} \mathrm{i} S^{y}$. Therefore, the Hamiltonian can be transformed into the Dicke representation as
\begin{equation}
  \begin{aligned}
    H=&-\frac{2\lambda}{N}\left[J_{N}^{2}-(J_{N}^{z})^{2}-\frac{N}{2}\right]-\left[S^{2}-(S^{z})^{2}-1\right]\\
    &-\frac{2\lambda}{N}\left(J_{N}^{-}S^{+}+J_{N}^{+}S^{-}\right)-2\cos{\theta}\left(J_{N}^{z}+S^{z}\right)\\
    &-2\sin{\theta}\left(J_{N}^{x}+S^{x}\right),
  \end{aligned}
  \label{1}
\end{equation}
where $J_{N}^{2}=\left(J_{N}^{x}\right)^{2}+\left(J_{N}^{y}\right)^{2}+\left(J_{N}^{z}\right)^{2}$ and $S^{2}=S_{x}^{2}+S_{y}^{2}+S_{z}^{2}$ characterize the total angular momentum operator of the bath and the radical pair, respectively.

In avian navigation, the spin-dependent chemical reaction only occurs in radical-pair system and depletes the radical pair to generate different chemical products according to their spin states, i.e., singlet or triplet state \cite{Ritz2000BJ,Yang2012PRA}. For the sake of simplicity, we assume that the singlet and triplet states have the same recombination constant $k$, and neglect the quantum jumps in the system. The time evolution of the density matrix of the total system is determined by \cite{Ritz2000BJ}
\begin{equation}
	\begin{aligned}
\frac{\mathrm{d}\hat{\rho}\left(t\right)}{\mathrm{d}t}=-\mathrm{i}\left[H,\hat{\rho}\left(t\right)\right]-k\hat{\rho}\left(t\right),
	\end{aligned}
	\label{3}
\end{equation}
where $k$ is the recombination rate.
The quantum yield of the singlet state in the chemical reaction can be calculated by using the projection operator $\hat{P}_{\mathrm{S}}^{\left(ab\right)}=\vert0,0\rangle_{a}~ _{b}\langle0,0\vert$ to the singlet state as
\begin{equation}
  \begin{aligned}
{\phi}_{\mathrm{S}}=k\int_{0}^{\infty} \mathrm{d}\tau\textrm{Tr}[\hat{P}_{\mathrm{S}}^{\left(ab\right)}\hat{\rho}\left(\tau\right)].
  \end{aligned}
  \label{2}
\end{equation}

Generally, the open quantum dynamics of the spin-dependent chemical reaction should be described by the quantum master equation \cite{Kominis2009PRE}. In Eq.~(\ref{3}), the quantum jump terms have been omitted and may lead to possible deviation from the experimental observation at low and high magnetic field due to quantum Zeno effect \cite{Kominis2009PRE,Ai2013SR}. Moveover, the exact quantum dynamics of the open quantum system can be given by the hierarchical equation of motion \cite{Ishizaki2009JCP}. Recently, it has been theoretically proposed and experimentally demonstrated that the exact solution can be exponentially accelerated by a quantum simulation approach \cite{Wang2018NPJQuanInf,Zhang2021FP}.

Assuming that initially the radical-pair system is in the singlet state, and the bath is in the thermal state $\sum_{M}\vert N/2,M \rangle\langle N/2,M\vert/(N+1)$, the initial density matrix is
\begin{equation}
  \begin{aligned}
    \hat{\rho}\left(t=0\right)=\frac{1}{N+1}\sum_{M=-N/2}^{N/2}\vert N/2,M \rangle\langle N/2,M\vert \otimes\hat{P}_{\mathrm{S}}^{\left(ab\right)}.
  \end{aligned}
  \label{4}
\end{equation}
Here, we use Dicke states $\left|N/2,M\right\rangle$ ($M=-N/2,\cdots, N/2)$ for the bath, and $\vert0,0\rangle,~\vert1,0\rangle,~\vert1,1\rangle, ~\vert1,-1\rangle$ for the radical pair.

It was suggested that birds' sensory transduction pathway can be effected by the quantum yield of the chemical reaction involved in radical-pair mechanism, and different amount of singlet-state quantum yield produces different visual modulation patterns, enabling birds to distinguish the direction of the geomagnetic field \cite{Ritz2000BJ}. Therefore, a reasonable model for avian navigation should have remarkable change in the quantum yield with respect to the variation of strength and direction of the magnetic field, so as to have a distinct sense of direction.

However, the quantum yield of this coupled system will remain a constant as the magnitude and direction of the magnetic field changes. This result can be expected by analysing the Hamiltonian~(\ref{1}). The operators of the radical pairs, $S$, $S^{z}$ and $S^{\pm}$ can not change the total spin number of the radical pair, making transition between the singlet and triplet states unavailable. Under this circumstance, no matter which initial state we choose for the radical bath, these two kinds of spin states will respectively convert to the corresponding chemical products without effecting each other.
Therefore, this multi-radical LMG model can not be used in the avian navigation. To solve this problem, we propose two models based on the LMG model for the avian navigation. The design principle of these two models is to break the symmetry between radical $a$ and $b$ in the Hamiltonian, in order to induce the transition between the singlet and triplet states of the radical-pair system.

\subsection{Absence of Zeeman Effect on Radical $b$}\label{Zeeman Effect}
The symmetry of the Hamiltonian can be broken by neglecting Zeeman Effect of the radical $b$, making the Hamiltonian of the system become
  \begin{align}
    H_{1}=&-\frac{\lambda}{N}\sum_{i<j}^{N}\left(\sigma_{i}^{x}\sigma_{j}^{x}+\sigma_{i}^{y}\sigma_{j}^{y}\right)-\frac{\lambda}{2}\left(\sigma_{a}^{x}\sigma_{b}^{x}+\sigma_{a}^{y}\sigma_{b}^{y}\right)\notag\\
    &-\frac{\lambda}{N}\sum_{i=1}^{N}\left[\sigma_{i}^{x}\left(\sigma_{a}^{x}+\sigma_{b}^{x}\right)+\sigma_{i}^{y}\left(\sigma_{a}^{y}+\sigma_{b}^{y}\right)\right]\notag\\
    &-\sum_{i=1}^{N}\left(\cos{\theta}\sigma_{i}^{z}+\sin{\theta}\sigma_{i}^{x}\right)-\left(\cos{\theta}\sigma_{a}^{z}+\sin{\theta}\sigma_{a}^{x}\right).
  \label{H1}
  \end{align}

To figure out the impact of this Hamiltonian on the quantum states, we transform the operators of single spin to those of multiple spins. Therefore, the Hamiltonian can be finally transformed into

\begin{equation}
  \begin{aligned}
    H_{1}=&-\frac{2\lambda}{N}\left[J_{N}^{2}-\left(J_{N}^{z}\right)^{2}-\frac{N}{2}\right]
    -\lambda\left[S^{2}-\left(S^{z}\right)^{2}-1\right]\\
    &-\frac{2\lambda}{N}\left(J_{N}^{-}S^{+}+J_{N}^{+}S^{-}\right)
     -2\cos{\theta}\left(J_{N}^{z}+S_{a}^{z}\right)\\
    &-2\sin{\theta}\left(J_{N}^{x}+S_{a}^{x}\right).
    \label{Jg}
  \end{aligned}
\end{equation}

Since there's no invariant subspace in the Hamiltonian with dimension $4(N+1)$. After careful calculation, the Hamiltonian is given as
\begin{equation}
\begin{aligned}
  &H_{1}=\\
  &\begin{bmatrix}
    \Lambda_{1}(-\frac{N}{2},\theta) & R_{1}^{\dagger}(-\frac{N}{2},\theta)         &\\
    R_{1}(-\frac{N}{2},\theta)       & \Lambda_{1}(-\frac{N}{2}+1,\theta) &\\
                              & R_{1}(-\frac{N}{2}+1,\theta)       & \ddots \\
                              & &\ddots &R_{1}^{\dagger}(\frac{N}{2}-2,\theta)\\
                              & & & \Lambda_{1}(\frac{N}{2}-1,\theta) & R_{1}^{\dagger}(\frac{N}{2}-1,\theta)\\
                              &                             &        & R_{1}(\frac{N}{2}-1,\theta) & \Lambda_{1}(\frac{N}{2},\theta)
    \end{bmatrix}
  \end{aligned}
  \label{HM}
\end{equation}
where $\Lambda_{1}(M)$ are the diagonal blocks and $R_{1}(M)$ are the nondiagonal blocks. The details are shown in Appendix~\ref{Appendix:IS}. For a specific case with $\theta=0$, the Hamiltonian $H$ is simplified as a block-diagonal matrix, cf. Appendix~\ref{section3}.

\subsection{Absence of Spin $b$'s Interaction with Bath} \label{Interaction}
We can also break the conservation of total spin number of the radical pair by neglecting the interaction between the bath and the radical $b$. Thus, the Hamiltonian becomes
\begin{equation}
  \begin{aligned}
    H_{2}=&-\frac{\lambda}{N}\sum_{i<j}^{N}\left(\sigma_{i}^{x}\sigma_{j}^{x}+\sigma_{i}^{y}\sigma_{j}^{y}\right)-\frac{\lambda}{2}\left(\sigma_{a}^{x}\sigma_{b}^{x}+\sigma_{a}^{y}\sigma_{b}^{y}\right)\\
    &-\frac{\lambda}{N}\sum_{i=1}^{N}\left(\sigma_{i}^{x}\sigma_{a}^{x}+\sigma_{i}^{y}\sigma_{a}^{y}\right)-\sum_{i=1}^{N}\left(\cos{\theta}\sigma_{i}^{z}+\sin{\theta}\sigma_{i}^{x}\right)\\
    &-\cos{\theta}\left(\sigma_{a}^{z}+\sigma_{b}^{z}\right)-\sin{\theta}\left(\sigma_{a}^{x}+\sigma_{b}^{x}\right).
  \end{aligned}
  \label{H2}
\end{equation}
After transformed into the Dicke representation, it reads
\begin{equation}
  \begin{aligned}
    H_{2}=&-\frac{2\lambda}{N}\left[J_{N}^{2}-(J_{N}^{z})^{2}-\frac{N}{2}\right]-\lambda\left[S^{2}-(S^{z})^{2}-1\right]\\
    &-\frac{2\lambda}{N}\left(J_{N}^{-}S_{a}^{+}+J_{N}^{+}S_{a}^{-}\right)-2\cos{\theta}\left(J_{N}^{z}+S^{z}\right)\\
    &-2\sin{\theta}\left(J_{N}^{x}+S^{x}\right),
  \end{aligned}
  \label{H22}
\end{equation}
where $S_{a}=\frac{1}{2}\sigma_{a}$. This system can be realized if the radical $b$ is further away from the bath than radical $a$, since the dipole-dipole interaction between the radicals decays with $1/r^{3}$, making the interaction sensitive to the distance. Considering the invariant subspaces in Eq.~(\ref{Jg}), we can also write this Hamiltonian as
\begin{equation}
\begin{aligned}
  &H_{2}=\\
  &\begin{bmatrix}
    \Lambda_{2}(-\frac{N}{2},\theta) & R_{2}^{\dagger}(-\frac{N}{2},\theta)         &\\
    R_{2}(-\frac{N}{2},\theta)       & \Lambda_{2}(-\frac{N}{2}+1,\theta) &\\
                              & R_{2}(-\frac{N}{2}+1,\theta)       & \ddots \\
                              & &\ddots &R_{2}^{\dagger}(\frac{N}{2}-2,\theta)\\
                              & & & \Lambda_{2}(\frac{N}{2}-1,\theta) & R_{2}^{\dagger}(\frac{N}{2}-1,\theta)\\
                              &                             &        & R_{2}(\frac{N}{2}-1,\theta) & \Lambda_{2}(\frac{N}{2},\theta)
    \end{bmatrix}
  \end{aligned}
  \label{H2M}
\end{equation}
where the explicit expressions of $\Lambda_{2}$ and $R_{2}$ are given in Appendix~\ref{Appendix:IS}.

\section{Numerical Simulation and Discussion}
\label{Sec:Simulation}

\begin{figure}[t]
	\includegraphics[width=8.8cm]{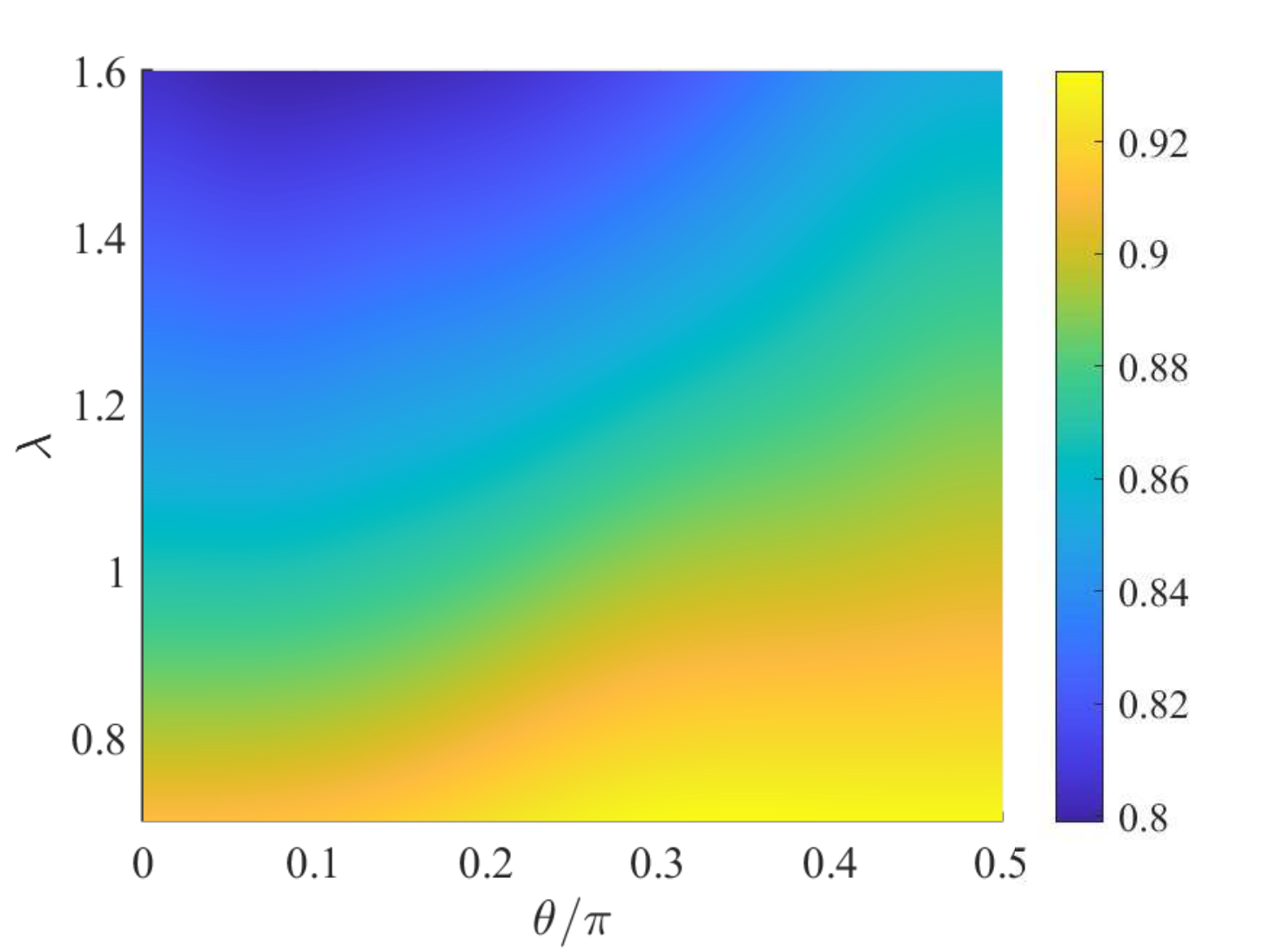}
	\caption{Quantum yield ${\phi}_{\mathrm{S}}$ varies with respect to the magnetic field's direction $\theta$ and strength $\lambda$ for $N=50$, $k=0.7~\mu$s$^{-1}$ based on Eq.~(\ref{H2}). The color in the figure represents the magnitude of production rate ${\phi}_{\mathrm{S}}$.}
	\label{fig:H2}
\end{figure}

\begin{figure}[t]
    \centering
    \includegraphics[width=8.8cm]{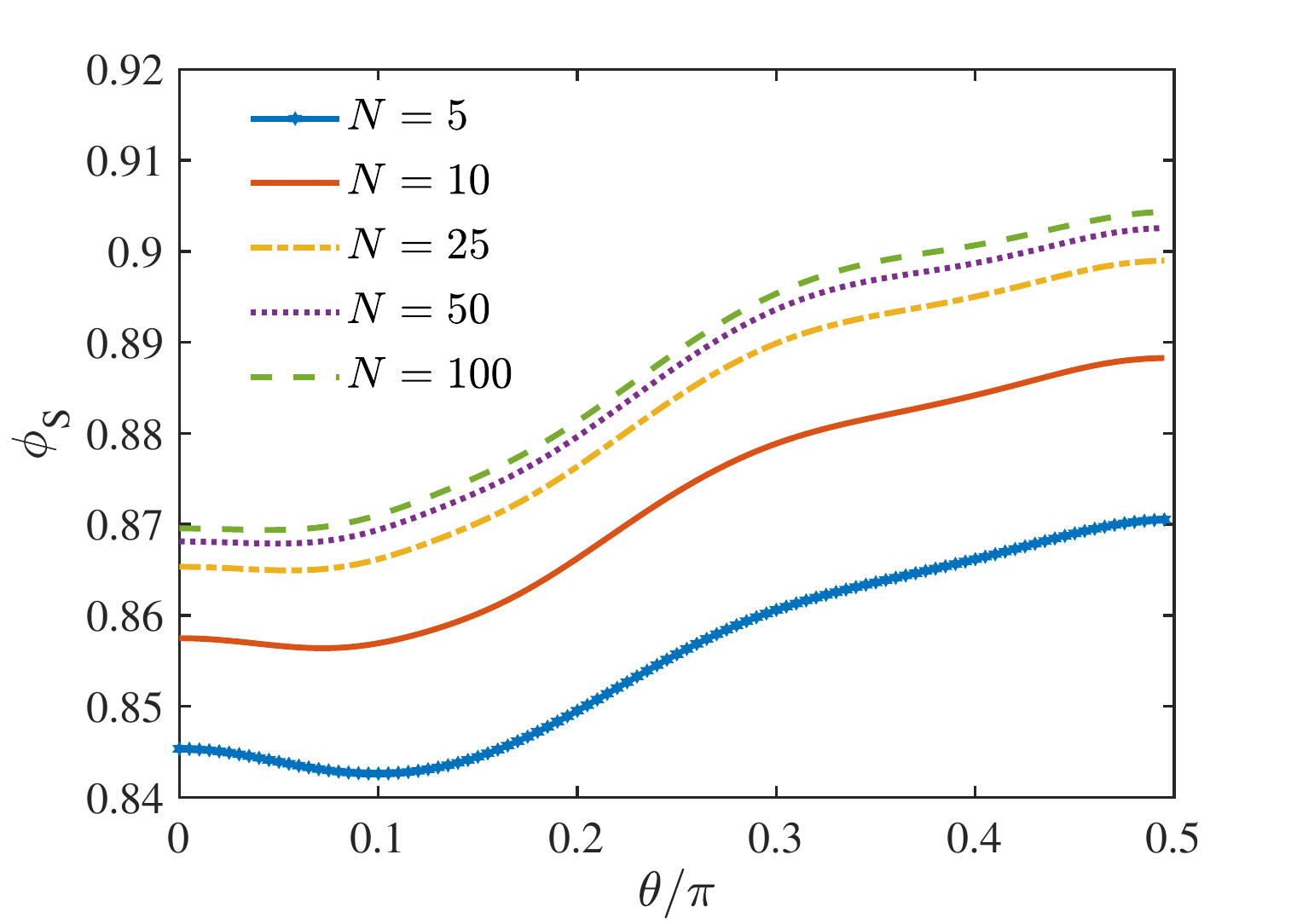}
    \caption{Quantum yield ${\phi}_{\mathrm{S}}$ varies with respect to the magnetic field's direction $\theta$ with $\lambda=1$ and $k=0.7~\mu$s$^{-1}$ for $N=5,10,25,50,100$, based on Eq.~(\ref{H2}). The quantum yield converges as $N$ increases.}
    \label{fig:H2k22}
\end{figure}

In Fig.~\ref{fig:H2}, our calculation shows that in the model for Eq.~(\ref{H2}), the quantum yield ${\phi}_{\mathrm{S}}$ of the singlet state will change remarkably with respect to both the direction and magnitude of the magnetic field, which covers a range from $0.8$ to $0.93$, making it feasible for avian navigation. We notice that the QPT takes place at $\lambda=1$, at which the quantum yield spans a wider range as $\theta$ changes. This implies that the sensitivity of avian navigation can be enhanced by the QPT. Furthermore, the QPT occurs only in the thermodynamic limit, i.e., $N\rightarrow\infty$. We also investigate the dependence of ${\phi}_{\mathrm{S}}$ on $N$ in Fig.~\ref{fig:H2k22}. As shown, the line will converge as $N$ increases. And the difference between the lines for $N=50$ and $N=100$ is small. Therefore, in Fig.~\ref{fig:H2}, we demonstrate the result for $N=50$.

\begin{figure}[t]
    \centering
    \includegraphics[width=8.8cm]{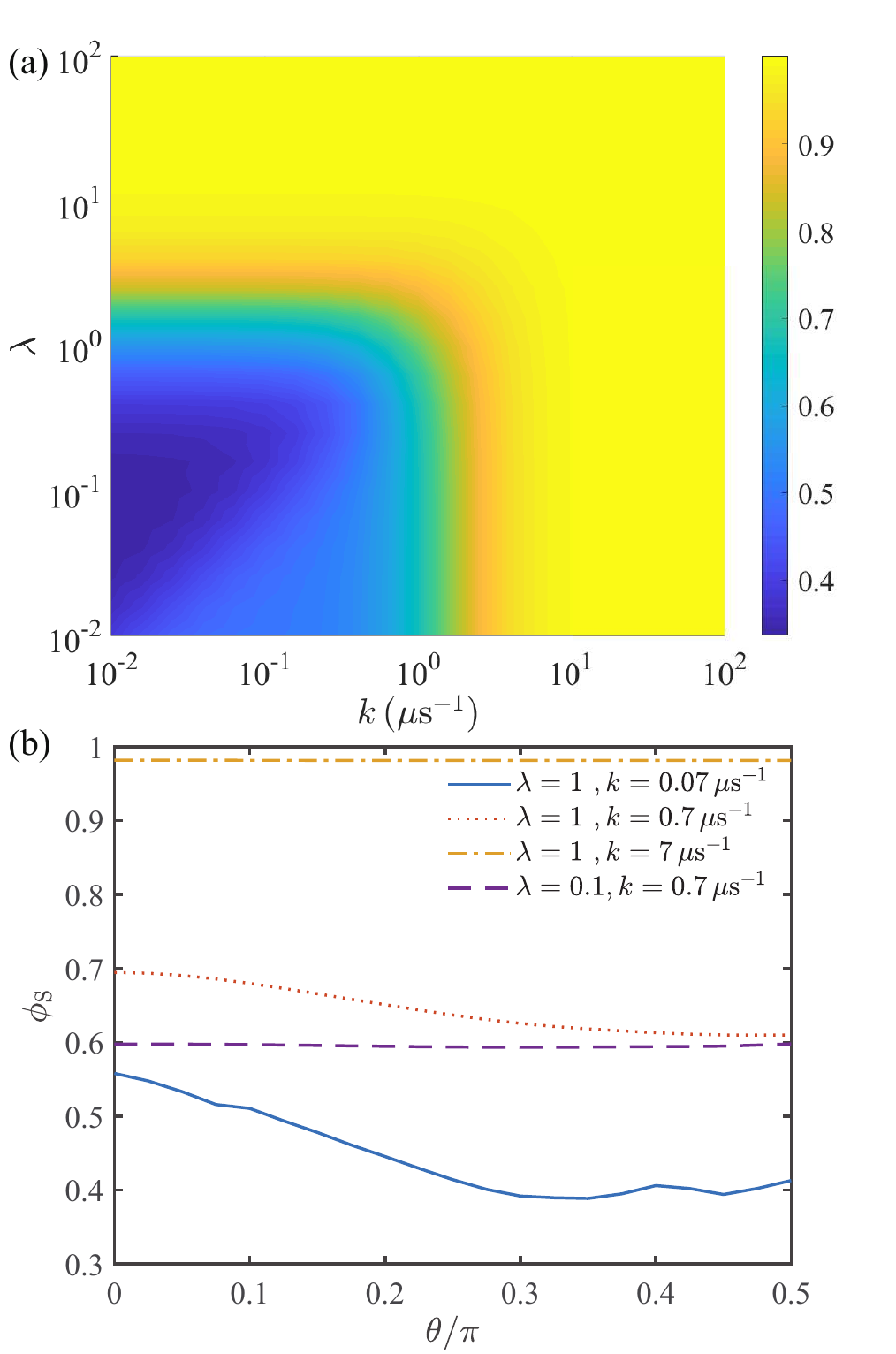}
    \caption{(a) Quantum yield ${\phi}_{\mathrm{S}}$ varies with respect to the magnetic field's strength $\lambda$ and recombination rate $k$ for $N=50$, based on Eq.~(\ref{H1}). (b) The sensitivity of avian navigation at different sets of $\lambda$ and $k$.}
    \label{fig:H1klambda}
\end{figure}

\begin{figure}[t]
    \centering
    \includegraphics[width=9cm]{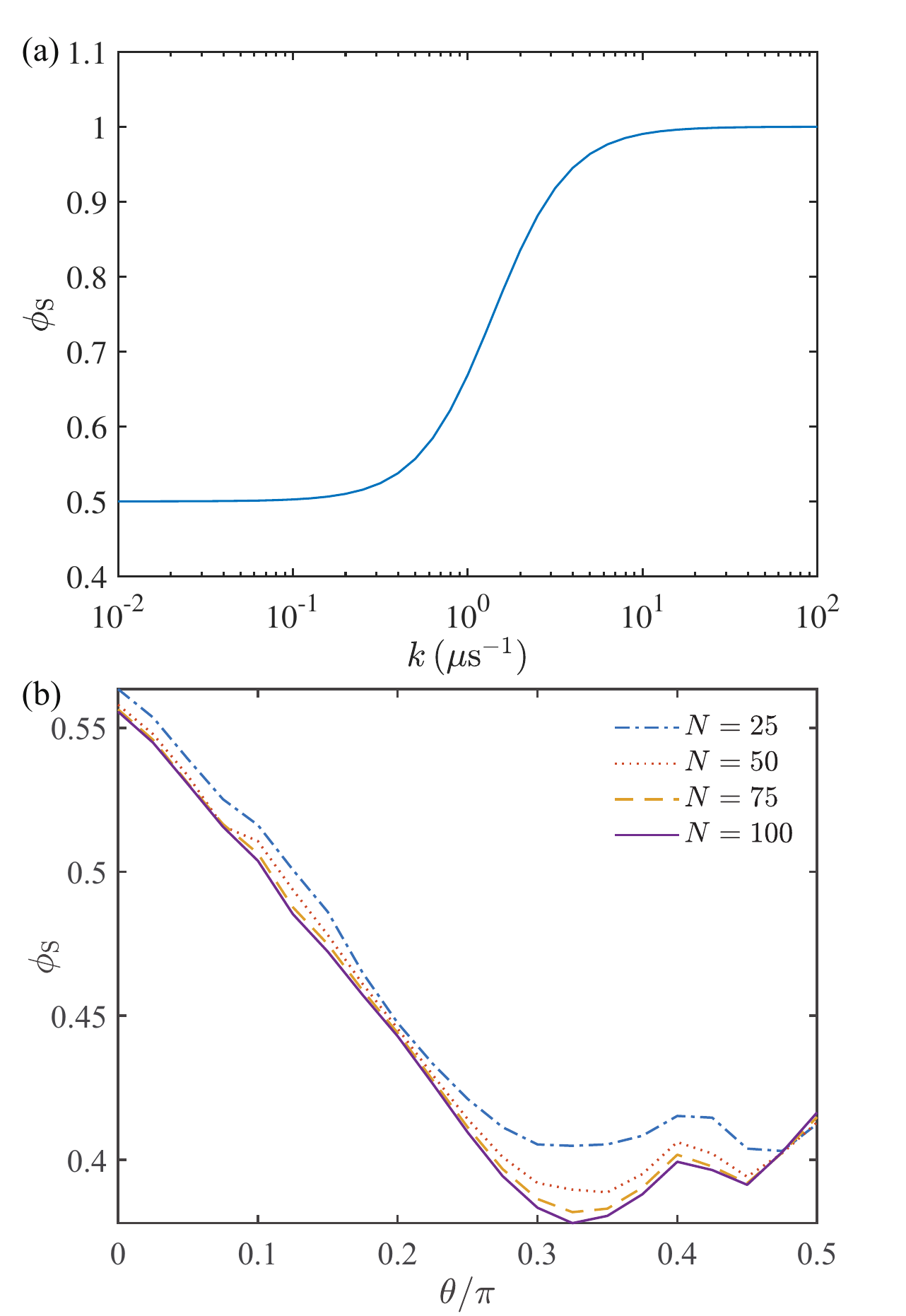}
    \caption{(a) The behavior of quantum yield $\phi_{\mathrm{S}}$ when $k$ goes to 0. This figure is obtained when $\lambda=0$ and $N=50$. As $k$ approaches 0, $\phi_{\mathrm{S}}$ approaches 0.5. (b) Quantum yield varies with the angle of the magnetic field $\theta$ at different $N$'s. Here we set $k=0.07\,\mathrm{\mu s}^{-1}$ and $\lambda=1$, which corresponds to the solid blue line in Fig.~\ref{fig:H1klambda}(b).}
    \label{fig:kphi}
\end{figure}

Hereafter, we will investigate the avian navigation for Eq.~(\ref{H1}). Figure~\ref{fig:H1klambda}(a) shows the dependence of $\phi_{\mathrm{S}}$ on $\lambda$ and $k$ for $\theta=0$ when there exist invariant subspaces, cf. Appendix~\ref{Appendix:IS}. At the first glance, there is a wide flat region in the outer parameter space, which is close to unity. When $k$ approaches infinity as $\lambda$ stays constant, there is no probability for the initial singlet state to convert to the triplet states and thus $\phi_{\mathrm{S}}\simeq1$ because the chemical reaction is too fast. On the contrary, when $k$ is close to 0, the radical pair will have sufficient time to evolve. Notice that for $\theta=0$ Eq.~(\ref{HM}) will become a block-diagonal matrix, cf. Appendix~\ref{Appendix:IS}. As an  approximation, we assume the diagonal terms to be identical and $\Lambda_{1}(M)$ in Eq.~(\ref{HM}) can be written as
\begin{equation}
    \Lambda_{1}(M)=
    \begin{bmatrix}
      \alpha & \lambda y(M) & 0 & 0\\
      \lambda y(M) & \alpha & \lambda x(M) & -1\\
      0 & \lambda x(M) & \alpha & 0\\
      0 & -1 & 0 &\alpha
    \end{bmatrix},
\end{equation}
whose eigenvectors are respectively
$(1/\lambda y,0,0,1)^{T}$,
$(-x/y,0,1,0)^{T}$,
$(-y\lambda,w,-x\lambda,1)^{T}$, and
$(-y\lambda,-w,-x\lambda,1)^{T}$, with $w=\sqrt{1+\lambda^{2}(x^{2}+y^{2})}$. As $\lambda$ goes to infinity, the first eigenvector approaches $(0,0,0,1)^{T}$, which is exactly the singlet state. This implies that the singlet state forms a invariant subspace and there is no conversion between the singlet and triplet states. Hence the quantum yield is almost 1 at $k\rightarrow0$ and $\lambda\rightarrow\infty$, which leads nearly-unity plateau in the top left part of Fig.~\ref{fig:H1klambda}(a). However, as $\lambda$ vanishes, the last two eigenvectors become $(0,1,0,1)^{T}$ and $(0,-1,0,1)^{T}$, respectively. In this case, each of them make up 50\% of the initial state. As $k>\lambda\rightarrow0$, the radical pair has sufficient time to oscillate between these two states, and thus there is $\phi_{\mathrm{S}}\simeq0.5$ at the lower triangle of the bottom left of Fig.~\ref{fig:kphi}(a). In order to check the validity of the above analysis, we numerically calculate the quantum yield as a function of chemical-reaction rate for $\lambda=0$ in Fig.~\ref{fig:kphi}(a). As $k$ decreases, $\phi_{\mathrm{S}}$ dramatically drops from unity to half around $k=1~\mu$s$^{-1}$.

In Fig.~\ref{fig:H1klambda}(b), we analyse the sensitivity for Eq.~(\ref{H1}) for four different sets of $\lambda$ and $k$. When the bath is far away from the QPT, e.g. $\lambda=0.1$, the quantum yield nearly stays constant for different $\theta$'s. When the bath is at the critical point, we significantly decrease $k$ from $7~\mu$s$^{-1}$ to $0.07~\mu$s$^{-1}$. When the chemical reaction happens too fast, the quantum yield still remains unchanged as the direction of magnetic field varies. When $k$ decreases, the quantum yield becomes lower than unity and shows its dependence on the direction. Thus, it improves sensitivity of the avian navigation.

According to Fig.~\ref{fig:H1klambda}(b), it seems that the QPT assists the avian navigation. In general, the true QPT occurs in the thermodynamic limit. As a result, we further investigate the sensitivity as $N$ is enlarged. When we constantly increase $N$, the results seem to asymptotically approach a limit, which implies our calculation is reliable. And the oscillation around $\theta\in[0.3\pi,0.5\pi]$ becomes more and more profound. Thus, we may safely arrive at the conclusion that the QPT can sensitively support the avian navigation.




\section{Conclusion}
\label{Conclusion}

In this paper, inspired by the multi-radical model \cite{Keens2018PRL}, we propose three multi-radical models with the radical bath described by the LMG model for artificial avian navigation. When the total system including the central radical pair and bath radicals are fully described by the LMG model, it can not be effectively used for navigation as it will not induce the coherent transition between the singlet and triplet states. However, if one of the central radical pair does not interact with the magnetic field or the radical bath, the capacity of navigation emerges. We further show that the sensitivity of navigation can be enhanced around the critical point and can be further improved as the number of radicals increases.

\begin{acknowledgments}
This work is supported by Beijing Natural Science Foundation under Grant No.~1202017 and the National Natural Science Foundation of China under Grant Nos.~11674033, 11505007.
\end{acknowledgments}

\appendix

\section{Magnetic Field Parallel to $z$-axis}\label{section3}
\begin{figure}[t]
	\includegraphics[width=8.8cm]{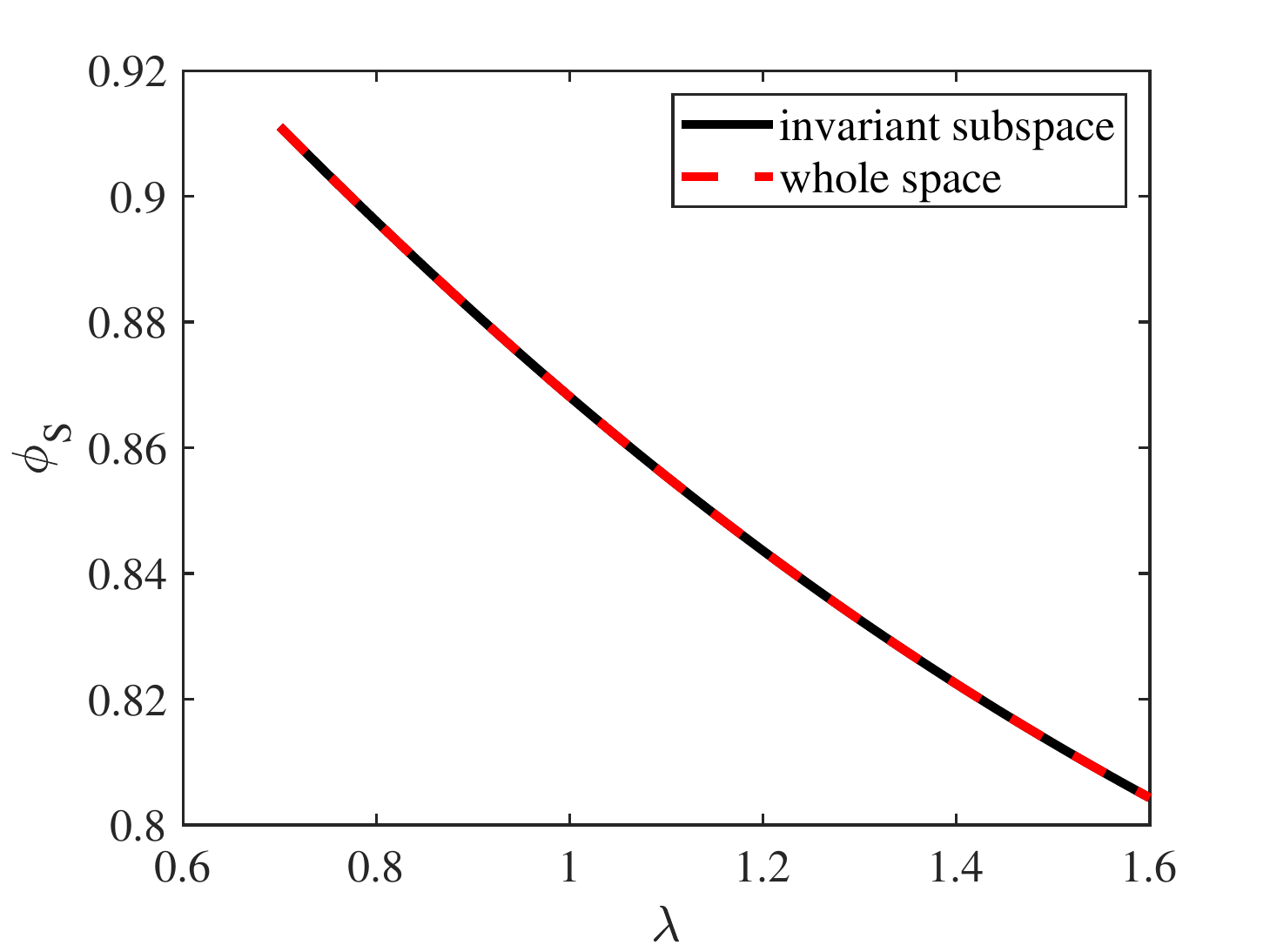}
	\caption{Comparison of the quantum yield ${\phi}_{\mathrm{S}}$ for using invariant subspaces and the whole space based on Eq.~(\ref{H2-1}) when $N=100$, $k=0.7\,\mathrm{{\mu}s}^{-1}$, $\theta=0$ . The black solid line represents the production rate based on invariant subspaces, and the blue dashed line represents the production rate based on the whole space.}
	\label{fig: comparation}
\end{figure}

In Section~\ref{Radical Pair Models}, we have discussed how the quantum yield of singlet state is influenced by the direction of the magnetic field. We use a $4(N+1)$-dimensional Hilbert space to calculate the quantum yield. However, when the magnetic field is parallel to $z$-axis and thus $\theta=0$, $H_1$ and $H_2$ become respectively
\begin{equation}
  \begin{aligned}
    H_{1}=&-\frac{2\lambda}{N}\left[J_{N}^{2}-\left(J_{N}^{z}\right)^{2}-N\right]-2\lambda\left[S^{2}-(S^{z})^{2}-1\right]\\
    &-\frac{2\lambda}{N}\left(J_{N}^{-}S^{+}+J_{N}^{+}S^{-}\right)
    -2\left(J_{N}^{z}+S_{a}^{z}\right),
    \label{H1-1}
  \end{aligned}
\end{equation}

\begin{equation}
  \begin{aligned}
    H_{2}=&-\frac{2\lambda}{N}\left[J_{N}^{2}-(J_{N}^{z})^{2}-\frac{N}{2}\right]-2\lambda\left[S^{2}-(S^{z})^{2}-1\right]\\
    &-\frac{2\lambda}{N}\left(J_{N}^{-}S_{a}^{+}+J_{N}^{+}S_{a}^{-}\right)-2\left(J_{N}^{z}+S^{z}\right).
  \end{aligned}
  \label{H2-1}
\end{equation}
As shown in Appendix~\ref{Appendix:IS}, we find that there exist invariant subspaces for both of these two models, which are spanned by
\begin{equation}
  \begin{aligned}
    \mathcal{H}=
    \{ & \left|N/2,M+1\right\rangle\left|1,-1\right\rangle,
	\left|N/2,M\right\rangle\left|1,0\right\rangle,\\
	& \left|N/2,M-1\right\rangle\left|1,1\right\rangle,
	\left|N/2,M\right\rangle\left|0,0\right\rangle\}.
  \end{aligned}
  \label{subspaces}
\end{equation}
Due to the existence of invariant subspaces, our calculations of the quantum  yield will be significantly simplified. Furthermore, by numerical simulations in Fig.~\ref{fig: comparation}, we show that the chemical yield of singlet state coincide with each other, and thus these two methods are the equivalent.

\section{Invariant Subspace}\label{Appendix:IS}
In this appendix, we will show that there exist invariant subspaces in the Hilbert space of Hamiltonians ~(\ref{H1-1}) and~(\ref{H2-1}) for the vertical situation, namely $\theta=0$. We can prove that the total angular momentum along $z$-axis, i.e., $J_{N}^{z}+S^{z}$, is conserved for both cases when $\theta=0$, by calculating the commutator with the corresponding Hamiltonian as
\begin{equation}
	\begin{aligned}
		\left[H_{1},J_{N}^{z}+S^{z}\right]
		=&-\left[2\sin\theta\left(J_{N}^{x}+S_{a}^{x}\right),\left(J_{N}^{z}+S^{z}\right)\right]\\
		=&\mathrm{i}2\sin\theta\left(J_{N}^{y}+S_{a}^{y}\right),\\
	\end{aligned}
	\label{Com2H1}
\end{equation}

\begin{equation}
	\begin{aligned}
		\left[H_{2},J_{N}^{z}+S^{z}\right]
		=&-\left[2\sin\theta\left(J_{N}^{x}+S^{x}\right),\left(J_{N}^{z}+S^{z}\right)\right]\\
		=&\mathrm{i}2\sin\theta\left(J_{N}^{y}+S^{y}\right).\\
	\end{aligned}
	\label{Com2H2}
\end{equation}

If $\theta\ne0$, the total angular momentum is not conserved along $z$-axis. The Hamiltonian $H_1$ can be rewritten as Eq.~(\ref{HM}), where
\begin{equation}
\begin{aligned}
  &R_{1}(M)=\\
  &\begin{bmatrix}
    -\sin\theta\zeta_{M+2} &-\frac{1}{\sqrt{2}}\sin{\theta} &0 &-\frac{1}{\sqrt{2}}\sin{\theta}\\
    0   &-\sin{\theta}\zeta_{M+1} &-\frac{1}{\sqrt{2}}\sin{\theta}  &0 \\
    0   &0    &-\sin{\theta}\zeta_{M} &0\\
    0   &0    & \frac{1}{\sqrt{2}}\sin{\theta}    &-\sin{\theta}\zeta_{M+1}
    \end{bmatrix}
  \end{aligned},
\end{equation}
\begin{equation}\label{Lambda1}
\begin{aligned}
  &\Lambda_{1}(M)=\\
  &\begin{bmatrix}
    d^{(1)}_{M}\left(\theta\right) & -\frac{2\sqrt{2}\lambda}{N}\zeta_{M+1} &0  &0\\
    -\frac{2\sqrt{2}\lambda}{N}\zeta_{M+1}   & b^{(1)}_{M}\left(\theta\right) &-\frac{2\sqrt{2}\lambda}{N}\zeta_{M}    &-\frac{2\lambda}{N}\cos\theta \\
    0  &-\frac{2\sqrt{2}\lambda}{N}\zeta_{M} &a^{(1)}_{M}\left(\theta\right) &0\\
    0  &-\frac{2\lambda}{N}\cos\theta     &0        &c^{(1)}_{M}\left(\theta\right)
    \end{bmatrix}
  \end{aligned}.
\end{equation}
Here,
\begin{subequations}
    \begin{align}
    \zeta_{M}&=\sqrt{\left(\frac{N}{2}-M+1\right)\left(\frac{N}{2}+M\right)},
    \label{zeta}\\
    a_{M}^{(1)}\left(\theta\right)&=-\frac{2\lambda}{N}\left[\frac{N^{2}}{4}-\left(M-1\right)^{2}\right]-\left(2M-1\right)\cos{\theta},\\
    b_{M}^{(1)}\left(\theta\right)&=-\frac{2\lambda}{N}\left(\frac{N^{2}}{4}-M^{2}\right)-\lambda-2M\cos{\theta},\\
    c_{M}^{(1)}\left(\theta\right)&=-\frac{2\lambda}{N}\left(\frac{N^{2}}{4}-M^{2}\right)+\lambda-2M\cos{\theta},\\
    d_{M}^{(1)}\left(\theta\right)&=-\frac{2\lambda}{N}\left[\frac{N^{2}}{4}-\left(M+1\right)^{2}\right]-\left(2M+1\right)\cos{\theta}.
    \end{align}
\end{subequations}

When the magnetic field is vertical to the plane, i.e., $\theta=0$, the commutator is $0$ and thus total angular momentum is conserved along $z$-axis. Therefore, the invariant subspaces in Eq.(\ref{subspaces}) will appear in those two Hamiltonians. We can rewrite $H_{1}$ as
\begin{equation}
\begin{aligned}
  H_{1}=
  \begin{bmatrix}
    \Lambda_{1}(-\frac{N}{2},0) &        &\\
                              & \ddots & \\
                              &        & \Lambda_{1}(\frac{N}{2},0)
    \end{bmatrix}
  \end{aligned}.
\end{equation}

$H_{2}$ is generally given as Eq.~(\ref{H2M}), where
\begin{equation}
\begin{aligned}
  R_{2}(M,\theta)=-\begin{bmatrix}
      \zeta_{M+2}\sin{\theta}   &  0  &  0  &  0\\
      \sqrt{2}\sin{\theta}  &  \zeta_{M+1}\sin{\theta}  &  0  &  0\\
      0  &  \sqrt{2}\sin{\theta}  &  \zeta_{M}\sin{\theta} &  0\\
      0  &  0                      &  0  &  \zeta_{M+1}\sin{\theta}
    \end{bmatrix}
  \end{aligned},
  \label{nondiagH2}
\end{equation}
\begin{equation}
\begin{aligned}
  \Lambda_{2}(M,\theta)=\begin{bmatrix}
      d_{M}^{(2)}(\theta)  &  -\frac{\lambda}{2\sqrt{2}}\zeta_{M+1}  &  0  &  -\frac{\lambda}{2\sqrt{2}}\zeta_{M+1}\\
      -\frac{\lambda}{2\sqrt{2}}\zeta_{M+1}  &  b_{M}^{(2)}(\theta)  &  0  &  -\frac{\lambda}{2\sqrt{2}}\zeta_{M}\\
      0  &  -\frac{\lambda}{2\sqrt{2}}\zeta_{M}  &  a_{M}^{(2)}(\theta) &  \frac{\lambda}{2\sqrt{2}}\zeta_{M}\\
      -\frac{\lambda}{2\sqrt{2}}\zeta_{M+1}  &  0  &  \frac{\lambda}{2\sqrt{2}}\zeta_{M}  &  c_{M}^{(2)}(\theta)
    \end{bmatrix}
  \end{aligned}.
\end{equation}
Here,
\begin{subequations}
\begin{align}
    a_{M}^{(2)}(\theta) &= a_{M}^{(1)}(\theta)+\cos{\theta},\\
    b_{M}^{(2)}(\theta) &= b_{M}^{(1)}(\theta),\\
    c_{M}^{(2)}(\theta) &= c_{M}^{(1)}(\theta),\\
    d_{M}^{(2)}(\theta) &= d_{M}^{(1)}(\theta)-\cos{\theta}.
\end{align}
\end{subequations}
When $\theta=0$, Hamiltonian $H_{2}$ can be also written in the block-diagonal form as
\begin{equation}
\begin{aligned}
  H_{2}=
  \begin{bmatrix}
    \Lambda_{2}(-\frac{N}{2},0) &        &\\
                              & \ddots & \\
                              &        & \Lambda_{2}(\frac{N}{2},0)
    \end{bmatrix}
  \end{aligned}.
\end{equation}
In the above derivation, we have used the properties of raising and lowering operators of the bath and radical pair, i.e.,
\begin{subequations}
\begin{align}
    &J_{N}^{+}\left|\frac{N}{2},M\right>=
    \sqrt{\left(\frac{N}{2}-M\right)\left(\frac{N}{2}+M+1\right)}\left|\frac{N}{2},M+1\right>,&\\
    &J_{N}^{-}\left|\frac{N}{2},M\right>=
    \sqrt{\left(\frac{N}{2}+M\right)\left(\frac{N}{2}-M+1\right)}\left|\frac{N}{2},M-1\right>,&\\
    &S^{+}\left|0,0\right>=S^{-}\left|0,0\right>=S^{+}\left|1,1\right>=S^{-}\left|1,-1\right>=0.&
\end{align}
\end{subequations}
Notice that $s_{a}^{z}$ can induce the transition between $\vert0,0\rangle$ and $\vert1,0\rangle$, while $\vert1,\pm1\rangle$ are the eigen states of $s_{a}^{z}$, i.e.,
\begin{subequations}
\begin{align}
    S_{a}^{z}\left|0,0\right>
    &=S_{a}^{z}\frac{1}{\sqrt{2}}\left(\left|\uparrow\downarrow\right>-\left|\downarrow\uparrow\right>\right)
    =\frac{1}{2}\left|1,0\right>,\\
    S_{a}^{z}\left|1,0\right>
    &=S_{a}^{z}\frac{1}{\sqrt{2}}\left(\left|\uparrow\downarrow\right>+\left|\downarrow\uparrow\right>\right)
    =\frac{1}{2}\left|0,0\right>,\\
    S_{a}^{z}\left|1,1\right>
    &=S_{a}^{z}\left|\uparrow\uparrow\right>
    =\frac{1}{2}\left|1,1\right>,\\
    S_{a}^{z}\left|1,-1\right>
    &=S_{a}^{z}\left|\downarrow\downarrow\right>
    =-\frac{1}{2}\left|1,-1\right>.
\end{align}
\end{subequations}

%
%

\bibliography{paper-1-ref}

\end{document}